\documentclass[11pt]{article}

\usepackage{multirow}
\usepackage{booktabs}
\usepackage{xcolor}
\usepackage{svg}
\usepackage{amsfonts}
\usepackage{amsmath}
\usepackage{placeins}
\newcommand{\method}{ReBOL} % change TBD to your final name

\newcommand{\lw}{LW-one-call}
\newcommand{\rankgpt}{LW-window}

\usepackage[preprint]{acl}

\usepackage{times}
\usepackage{latexsym}

\usepackage[T1]{fontenc}
\usepackage[utf8]{inputenc}

\usepackage{microtype}

\usepackage{inconsolata}

\usepackage{graphicx}

\title{ReBOL: Retrieval via Bayesian Optimization with Batched LLM Relevance Observations and Query Reformulation}

\author{Anton Korikov \\
 University of Toronto  \\
  \texttt{korikov@mie.utoronto.ca} \\\And
  Scott Sanner \\
  University of Toronto\\
  \\}

\begin{document}
\maketitle
\begin{abstract}

LLM-reranking is limited by the top-$k$ documents retrieved by vector similarity, which neither enables contextual query-document token interactions nor captures multimodal relevance distributions. While LLM query reformulation attempts to improve recall by generating improved or additional queries, it is still followed by vector similarity retrieval.
%weak similarity scoring
We thus propose to address these top-$k$ retrieval stage failures by introducing 
\method, which 1) uses LLM query reformulations to initialize a multimodal Bayesian Optimization (BO) posterior over document relevance, and 2) iteratively acquires document batches for LLM query-document relevance scoring followed by posterior updates to optimize relevance. After exploring query reformulation and document batch diversification techniques, we evaluate \method{} against LLM reranker baselines on five BEIR datasets and using two LLMs (Gemini-2.5-Flash-Lite, GPT-5.2). \method{} consistently achieves higher recall and competitive rankings, for example compared to the best LLM reranker on the Robust04 dataset with 46.5\% vs. 35.0\% recall@100 and 63.6\% vs. 61.2\% NDCG@10. We also show that \method{} can achieve comparable latency to LLM rerankers.

\end{abstract}

%%%OLD ABSTRACT%%%%%
% LLM-augmented retrieval features two key paradigms: LLM query reformulation, which augments an initial query, and LLM reranking, where an LLM judges the relevance of a list of top-$k$ retrieved documents. However, both paradigms are typically bounded by the downstream/upstream performance of vector similarity retrieval, which encodes queries and documents separately and uses only a simple function such as dot product to evaluate query-document.  
%We thus ask whether LLM relevance signals from query reformulation and document relevance judgments can enable more expressive and higher recall retrieval from the full corpus, observing that searching a large document collection with expensive relevance observations naturally suggests a Bayesian optimization approach --- leading us to introduce \method{}, for \textbf{Re}trival via \textbf{B}ayesian \textbf{O}ptimization with \textbf{L}LM Relevance Observations.
% We evaluate \method{} against LLM reranker baselines with and without LLM query reformulation on five BEIR datasets and using two LLMs (Gemini-2.5-Flash-Lite, GPT-5.2). We find that \method{} typically results in much higher recall and competitive rankings, for example compared to the best LLM reranker on the Robust04 dataset with 46.5\% vs. 35.0\% recall@100 and 63.6\% vs. 61.2\% NDCG@10. We also show that \method{} can achieve comparable latency to LLM rerankers and investigate several diversified document batching techniques to improve efficiency and exploration. 

\section{Introduction}
\begin{figure*}[t]
    \centering
    \includegraphics[width=\linewidth]{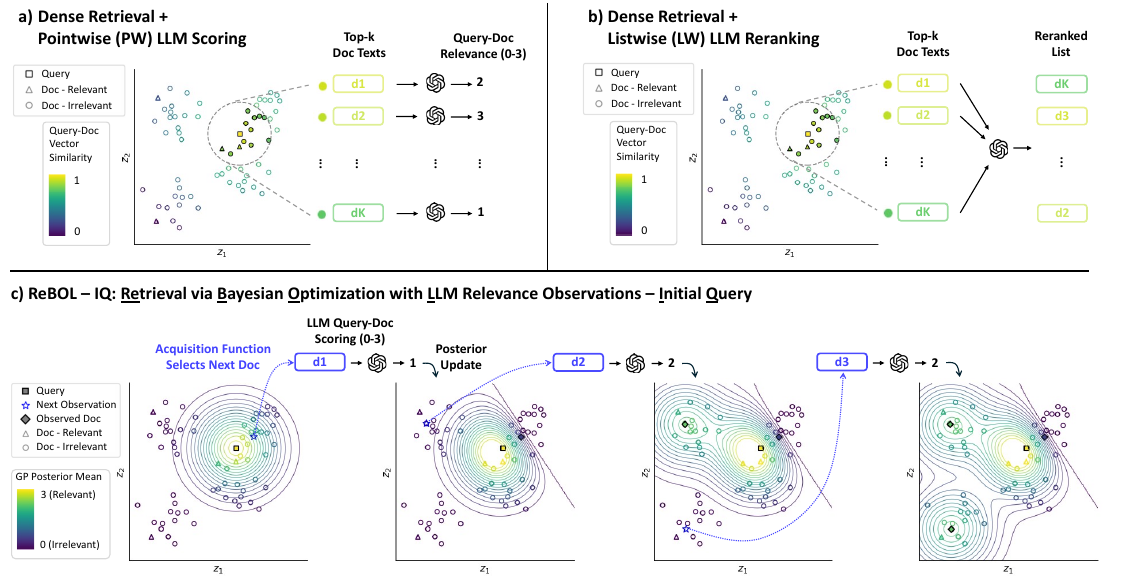}
    \caption{a-b) LLM rankers rely on upstream vector similarity retrieval to reduce a large document collection to a top-$k$ candidate list. c) \method-IQ (\textbf{I}nitial \textbf{Q}uery) first initializes a GP prior with a high-relevance point at the query embedding (yellow square), then iteratively selects which document to judge next using an acquisition function (blue star), scores its relevance to the query using an LLM, and updates a multimodal GP posterior.} 
    
    %uses LLM generated $q$-$d$ relevance scores (diamonds) and a multimodal Gaussian process to maintain a Bayesian belief in $q$-$d$ relevance over all of $\mathcal{X}$ and actively select which document to judge next using an acquisition function (blue star). The method is initialized (left) with a prior assuming low (purple) relevance (as most documents are likely irrelevant) plus the initial query embedding as the first observation which is assigned maximum relevance (yellow square).} 
    \label{fig:gp}
\end{figure*}

\begin{figure*}[t]
    \centering
    \includegraphics[width=\linewidth]{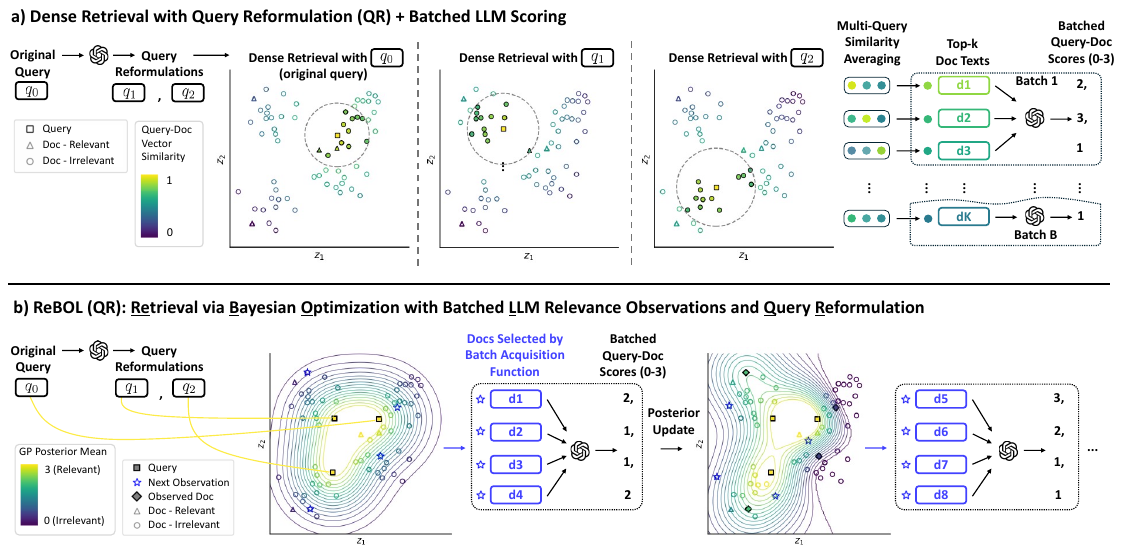}
    \caption{ 
    Given $q_0$, an LLM generates query reformulations $q_1$ and $q_2$. a) An example reformulate-retrieve-rerank pipeline, using each $q \in \{q_0,q_1,q_2\}$ for vector similarity retrieval before score averaging and batched top-$k$ LLM scoring 
    %(which can reduce latency and allow document interactions). 
    b) \method{}-QR (\textbf{Q}uery \textbf{R}eformulation) initializes a multimodal GP posterior with peaks at query embeddings $\{z^{q_0},z^{q_1},$ $z^{q_2}\}$ followed by active document batch acquisition, batched LLM scoring, and posterior updating.} 
    %--- providing a unified way to use $q_0$, $q_1,$ and $q_2$, and LLM $q$-$d$ relevance scores to maintain relevance beliefs --- also using batch acquisition functions for improved efficiency and exploration.}
    \label{fig:qr}
\end{figure*}

Given a query $q_0$, LLM rerankers (e.g., \citet{ma2023zero, upadhyay2024umbrela}) are limited to the top-$k$ vector similarity retrieved documents and cannot recover from low-recall retrieval. While a line of LLM query reformulation work (e.g., \citet{wang2023query2doc, kostric2024surprisingly}) aims to improve recall by generating one or more query reformulations $\{q_1,...,q_Q\}$, these reformulations are ultimately still processed by vector similarity retrieval. Unfortunately, vector similarity is limited to scoring separately encoded query and document vectors, precluding contextual query-document token interaction. Further, its use of simple scoring functions such as the dot product induces a unimodal relevance distribution around the query, while relevance is often distributed multi-modally across distinct document clusters \cite{van2004geometry}.

To overcome these failure modes of the top-$k$ retrieval stage of reformulate-retrieve--rerank pipelines, we instead consider interleaving retrieval with LLM relevance observations in a Bayesian optimization (BO) framework. First, BO enables us to use the initial query $q_0$ and its LLM reformulations $\{q_1,...,q_Q\}$ to initialize multiple relevance peaks in a multimodal posterior over document relevance (Fig. \ref{fig:qr}c, left). BO then lets us systematically select batches of documents for which to generate high-fidelity LLM relevance labels to update the posterior and optimize relevance (Fig. \ref{fig:qr}). 
%This LLM-BO retrieval framework aims to overcome the limitations of vector similariy top-$k$ retrieval. 
Two key questions that emerge are (a) how to initialize and seed multimodal BO using LLM query reformulations and (b) how to efficiently acquire informative document batches for LLM labeling and posterior updates. Thus,  in this work, we: 
\begin{itemize}
    \item Introduce \method{} (\textbf{Re}trival via \textbf{B}ayesian \textbf{O}ptimization with Batched \textbf{L}LM Relevance Observations) which initializes a multimodal BO posterior with LLM query reformulations and actively selects document batches for LLM labeling and relevance optimization.
    \item Explore efficient BO batch acquisition functions, introducing a novel, MMR \cite{carbonell1998use} inspired batch diversification method %(Sec. \ref{sec:meth-af}). 
    \item Evaluate \method{} on five BEIR \cite{thakur2021beir} datasets against  sparse retrieval baselines and LLM rerankers (with/without LLM query reformulation), using two LLMs (Gemini-2.5-Flash-Lite, GPT-5.2) as well as a cross-encoder.
    \item Demonstrate that \method{} consistently attains much higher recall and competitive rankings, for instance compared to the best LLM reranker (with LLM query reformulation) on the Robust04 dataset %\cite{voorhees2005overview}
    where it achieves 46.5\% recall@100 vs. 35.0\% and an NDCG@10 of 63.6\% vs. 61.2\%, respectively. %(Tab. \ref{tab:main}).
    \item Show that \method{} has comparable latency to LLM rerankers, since LLM latency generally dominates the extra BO overhead.% (Tab. [\textbf{REF}]).  
\end{itemize}

\section{Background and Related Work}\label{sec:background}
 
\subsection{Vector Similarity Retrieval}
Vector similarity retrieval relies on an encoder $g(x) = z^x$ which maps some text $x$ to a vector $z^x \in \mathbb{R}^m$. Documents are encoded offline, and given query $q$, the retriever computes $g(q)=z^q$ and scores it against each $z^d$
%$\in \{z^d\}_{d\in\mathcal{X}}$ 
using a similarity function 
$S(\cdot,\cdot):\mathbb{R}^m\times\mathbb{R}^m \rightarrow \mathbb{R} $
such as the dot product. Sparse retrievers such as BM25 \cite{robertson2009probabilistic} use token frequency vectors for lexical matching \cite{salton1975vector}, while dense retrievers (e.g., \citet{izacard2021unsupervised, gao2021condenser}) use neural encoders to produce dense embeddings that aim to capture semantic similarity.
%While vector similarity retrieval has very low latency due to independent encoding of the query and documents, it relies only on a simple vector similarity function to compute $q$-$d$ relevance --- thus imposing severe performance bounds on any downstream reasoning when used to reduce $\mathcal{X}$ to a top-$k$ candidate set.

\subsection{LLM-augmented Retrieval}
\subsubsection{LLM Query Reformulation} 
Given an initial query $q_0$, recent work shows that retrieval can be improved by using an LLM to generate a query reformulation (e.g., \citet{wang2023query2doc, jagerman2023query, dhole2024genqrensemble, wen2025simple}) or multiple reformulations $\{q_1,...,q_Q\}$ (e.g., \cite{lin2021multi, kostric2024surprisingly, dhole2024generative, korikov2024multiaspect})  to be then used for vector similarity retrieval. Multi-query retrieval results are typically aggregated via similarity averaging (e.g., Fig \ref{fig:qr}a) or ranked list fusion.

%Ultimately, these methods use generative LLMs to make better or more query vectors $\{z^{q_1},...,z^{q_Q}\}$ for similarity scoring.

\subsubsection{LLM Relative Relevance Judgment}\label{sec:back-lw}
Given $q_0$ and a top-$k$ document list, listwise (LW) and pairwise rerankers 
%prompt an LLM to 
judge relative relevance between documents. Specifically, LW methods prompt an LLM
%with a list of document texts and IDs to generate a new list of IDs in order of 
to reorder a document list by descending relevance to $q_0$ (e.g., Fig \ref{fig:gp}b). The most basic LW variation \cite{ma2023zero} processes all $k$ documents in a single LLM call, but including too many candidates can degrade performance, so stronger methods use an upwards sliding window strategy \cite{ma2023zero,sun2023chatgpt, pradeep2023rankvicuna, pradeep2023rankzephyr}. Pairwise LLM rerankers \cite{qin2024large,liu2024lost} limit the documents per LLM relative judgment call even further to only two, often performing strongly  --- but unfortunately require a quadratic number of LLM calls so are not included in our experiments.

%Generally, neither LW nor pairwise rerankers generate absolute relevance judgments.

\subsubsection{LLM Absolute Relevance Scoring}\label{sec:back-pw}
Generating real-valued LLM query-document relevance scores $s_{q,d} \in \mathbb{R}$ is often called pointwise (PW) reranking, and is typically done using some rubric such as the 0-3 UMBRELA labels below \cite{upadhyay2024umbrela}: 
%called pointwise
%generate scores $s_{q,d} \in \mathbb{R}$ 
%can use logits, batching
%scoring rubric can be semantic$
%example (used in this paper), UMBRELA
%cross-encoders? %\cite{nogueira2019passage, zhuang2023rankt5} which jointly embed queries and passages to predict a relevance score.
\begin{itemize}
    \item \textbf{3:} The passage is dedicated to the query and contains the exact answer.
    \item \textbf{2:} The passage has some answer for the query, but the answer may be a bit unclear, or hidden amongst extraneous information.
    \item \textbf{1:} The passage seems related to the query but does not answer it.
    \item \textbf{0:} The passage has nothing to do with the query.
\end{itemize}
Documents can be LLM-scored one-by-one \cite{sachan2022improving,upadhyay2024umbrela,tornberg2024best}
%\cite{sachan2022improving,zhuang2023beyond,thomas2024large,upadhyay2024umbrela,tornberg2024best, ma2024fine} 
or as a batch \cite{korikov2025batched}, with the latter often increasing efficiency and sometimes performance as it enables interactions between documents. %A number variations exist, including using score logits to compute continuous scores and self-consistency. 

\subsection{Bayesian Optimization}\label{sec:back-bo}
In BO \cite{garnett2023bayesian}, we start with a real-valued function over some domain $\mathcal{X}$; $f:\mathcal{X}\rightarrow \mathbb{R}$ which we assume is distributed according to some prior $p(f(x)|x)$, and our goal is to systematically search for $x^* \in \arg \max_{x \in \mathcal{X}} f(x)$. To gain information about $f$, which may be a black-box function, we have some mechanism to observe a $y \in \mathbb{R}$ at an arbitrary point $x$. We assume that $y$ is distributed according to an observation model $y 
\sim p(y|x,f(x))$, with a common observation model being Gaussian noise about an $f(x)$ mean:
\begin{equation} \label{eqn:gaus-obs}
    p(y|x,f(x),\sigma_n) = \mathcal{N}(y;f(x),\sigma_{n}).
\end{equation}
We also assume conditional independence between observations $\mathbf{y} = [y^1,...,y^t]$ at $\mathbf{x} = [x^1,...,x^t]$:
\begin{equation}
p(\mathbf{y}|\mathbf{x}, f(\mathbf{x})) = \prod_i^tp(y^i|x^i,f(x^i)),
\end{equation}
and Bayes theorem defines our new posterior as $p(f(\mathbf{x})|\mathcal{D}^t) \propto p(f(\mathbf{x})|\mathbf{x})p(\mathbf{y}|\mathbf{x}, f(\mathbf{x}))$ where $\mathcal{D}^t = (\mathbf{x}, \mathbf{y})$. At any time $t$, the posterior can be used to predict $f$ (e.g., Sec. \ref{sec:gp-back}, Eq \eqref{eqn-gp-pred-pos-mean},\eqref{eqn-gp-pred-pos-var}) and inform an acquisition function $\alpha:\mathcal{X}\rightarrow\mathbb{R}$ to select the next observation location $x^{t+1}$ as $\arg \max_x \alpha(x|\mathcal{D}^t)$ (Sec. \ref{sec:meth-af}).

% letting us use observations to update our beliefs over $f(\mathbf{x})$ via Bayes theorem to get the posterior $p(f(\mathbf{x})|\mathbf{x},\mathbf{y}) \propto p(f(\mathbf{x})|\mathbf{x})p(\mathbf{y}|\mathbf{x}, f(\mathbf{x})$.

% Multiple observations $
% are assumed conditionally independent so that $p(\mathbf{y}|\mathbf{x}, f(\mathbf{x})) = \prod_ip(y_i|x_i,f(x_i))$, letting us perform posterior updates incrementally. 

\subsubsection{Gaussian Processes}\label{sec:gp-back}
Since it enables efficient closed-form inference, a very common prior is the Gaussian process (GP), which for a zero mean is defined as $p(f) = GP(0,k)$, where $k:\mathcal{X} \times \mathcal{X} \rightarrow \mathbb{R}_0^+$ is called the kernel. For any two points, $k(x,x')$ represents the covariance between $f(x)$ and $f(x')$, and a common choice is the RBF kernel
\begin{equation}\label{eqn-rbf}
    k(x,x') = \sigma_s^2 \exp\!\left(-\frac{\|x-x'\|^2}{2\ell^2}\right),
\end{equation}
where $\ell$ is the length-scale controlling how quickly correlations decay, and $\sigma_s^2$ is the signal variance.

Given $t$ observations $\mathbf{x},\mathbf{y}$ following the Gaussian noise model in Eq. \eqref{eqn:gaus-obs}, the GP predictive posterior used to predict $f(x')$ at any $x' \in \mathcal{X}$ is $p(f(x')|\mathbf{x},\mathbf{y},x') = \mathcal{N}(\mu[f(x')],\sigma^2[f(x')])$ with:

\begin{align}
\mu [f(x')] &= \mathbf{k}^T[\mathbf{K}+\sigma_n^2 I]^{-1}\mathbf{y}, \label{eqn-gp-pred-pos-mean}\\
\sigma^2 [f(x')] &= k(x',x') - \mathbf{k}^T[\mathbf{K}+\sigma_n^2 I]^{-1}\mathbf{k}, \label{eqn-gp-pred-pos-var}
\end{align}
where $\mathbf{K}_{ij} = k(x_i, x_j)$, and $\mathbf{k}_i = k(x',x_i)$ for $i,j \in [1,...,t]$.

%\item Mention a few LLM-BO hybrid references

% \item Ability to get absolute $q-d$ relevance judgments is new. CE give score, and PW gives semantically meaning full scores (eg 2 is partially relevant, 3 is relevant).

\section{Methodology}\label{sec:meth}
To address the top-$k$ performance limits imposed by vector similarity retrieval on LLM query reformulation and reranking, we re-frame retrieval as a BO task, introducing \method{} 
(Fig. \ref{fig:gp}c \& \ref{fig:qr}b) --- which first uses 
$q_0$ and any LLM reformulations $\{q_1,...,q_Q \}$
%the initial and LLM generated query (reformulations) 
as observations to initiate a multimodal GP posterior (Sec. \ref{sec:rel-func}-\ref{sec:meth-qr}), followed by iterative selection of documents with a batch acquisition function (Sec. \ref{sec:meth-af}), batched LLM query-document relevance scoring (Sec. \ref{sec:meth-llm-score}), and posterior updating (Sec. \ref{meth:rel-pred}). 

\subsection{Query-Document Relevance Function}\label{sec:rel-func}
%\subsection{Relevance Function over Dense Embeddings}
We first define a query-document relevance function $f:\mathbb{R}^m\rightarrow\mathbb{R}$ over the embedding space of an $m$-dimensional dense text encoder $g(x) = z^x$, and let the set of all embedded documents $d \in \mathcal{X}$ be represented by $\mathcal{Z}^\mathcal{X} \subset \mathbb{R}^m$. While 
%not strictly neccessary, in this work we find it convinent to bound $f$ 
$f(z)$ could have any form, in this work we find it convenient to bound $f(z) \in [0,s^{\text{max}}]$ where 0 represents irrelevance and $s^{\text{max}} \in \mathbb{R}$ represents maximum relevance (e.g., 3 in the prompt in Sec. \ref{sec:back-pw}).
%IF NO SPACE: move multimodality arg to next par 
Unlike vector similarity, we assume $f$ can be multimodal, reflecting that relevance is often distributed between multiple clusters \cite{van2004geometry}.

\subsection{Multimodal GP Relevance Prior}
To enable efficient inference, we use a GP prior (Sec \ref{sec:gp-back}) over $f(z)$, using a zero mean to model the assumption that most documents are irrelevant for a given $q_0$. We use a non-linear kernel (e.g. RBF, Eq. \eqref{eqn-rbf}) to enable multimodality (e.g., see Figs. \ref{fig:gp}c \& \ref{fig:qr}b showing distinct relevance peaks).
%could move unimodal comment to footnote to emphasize it
In contrast, vector retrieval with dot product or (normalized) cosine similarity induces unimodal relevance due to its linear scoring function. 

%Two observation modes -- generating text that we assume should have a high relevance score (query reformulations), and predicting relevance scores for existing text (documents) selected by an acquisition function 

\subsection{Initial Query and LLM Reformulations}\label{sec:meth-qr}
The initial query $q_0$ provides a natural first observation since it is reasonable to assume maximal relevance at the query embedding, giving the observation pair $(z^1,y^1) = (z^{q_0}, s^{\text{max}})$, as shown on the left of Fig. \ref{fig:gp}c. Further, LLM-generated reformulations $\{q_1,\ldots,q_Q\}$ can capture different aspects of the query and be incorporated as analogous observations $(z^{i+1},y^{i+1}) = (z^{q_i}, s^{\text{max}})$ for $i \in [1,\ldots,Q]$, with the goal of inducing multiple GP relevance peaks in the embedding space. Our experiments consider two \method{} variants: \method-IQ, which uses only the initial query observation, and \method-QR, which incorporates query reformulations.

\subsection{LLM Document Relevance Scores}\label{sec:meth-llm-score}
After initialization with query observations, \method{} iteratively uses an acquisition function (Sec. \ref{sec:meth-af}) to select a batch of $B$ documents $\mathbf{d}^t = [d^t,...,d^{t+B-1}]$ at each time step $t$. An LLM is then used (as per Sec. \ref{sec:back-pw}) to generate relevance scores $\{s_{q_0,d^{t'}}\}$ with $t' \in [t,...,t+B-1]$, giving observations at document embeddings $(z^{t'},y^{t'}) = (z^{d^{t'}}, s_{q_0,d^{t'}})$. 
%If no space, move next point to AF section
Here, batched BO acquisition and LLM scoring can reduce latency and enable the potential benefits of inter-document interactions during LLM scoring \cite{korikov2025batched}, as well as provide an additional exploration mechanism via batch diversification, as discussed next. 

\subsection{Document  Acquisition Functions}\label{sec:meth-af}
While there exists a wide range of GP acquisition functions \cite{garnett2023bayesian}, we explore the following variants, including a novel MMR-style document batch diversification method.

\subsubsection{Single Document Acquisition}
% Recall from Section \ref{sec:back-bo} that a (single-point) acquisition function $\alpha:\mathcal{X}\rightarrow\mathbb{R}$ is used to select the next observation location $x^{t+1}$ as $\arg \max_x \alpha(x|\mathcal{D}^t)$ given observations $\mathcal{D}^t$. 
As per Sec. \ref{sec:back-bo}, we define a single-document acquisition function
$\alpha:\mathcal{Z}^{\mathcal{X}} \rightarrow \mathbb{R}$ over document embeddings $\mathcal{Z}^{\mathcal{X}}$, which is used to select the next document  $d^{t+1}$ as $\arg \max_{d \in \mathcal{X}} \alpha(z^d|\mathcal{D}^t)$ given observations $\mathcal{D}^t$. 

%if need space -- chunk this section into one or two pars.
\paragraph{Greedy} Our first acquisition function is \textit{greedy}, which simply equals the posterior mean (Eq. \eqref{eqn-gp-pred-pos-mean}), giving $\alpha(z^d|\mathcal{D}^t) = \mu [f(z^d)]$. Greedy is an exploitation-only strategy, meaning that it will prioritize observing documents that are embedded near the query, query reformulations, and any documents with high observed relevance. 

\paragraph{UCB} The Upper Confidence Bound (UCB) acquisition function balances exploration and exploitation, combining both the posterior mean (Eq. \eqref{eqn-gp-pred-pos-mean}) and uncertainty (Eq. \eqref{eqn-gp-pred-pos-var}) as 
\begin{equation}
\alpha(z^d|\mathcal{D}^t) = \mu[f(z^d)] + \sqrt{\beta}\,\sigma[f(z^d)],
\end{equation}
encouraging exploration of uncertain regions that may contain highly relevant documents, with $\beta$ controlling the exploration-exploitation tradeoff.%greedy -- could be OK since dense retriever give good q-d sim... 

\paragraph{Random} Tested as a maximal exploration baseline, this function selects documents randomly. 

\subsubsection{Document Batch Acquisition}\label{meth:batch-af}
To select a batch $\mathcal{B}^t$ of 
%of $B$ 
documents $\mathbf{d}^{t} = [d^{t},\ldots,d^{t+B-1}]$ at time step $t$, and given some single-point acquisition function, we explore the following (diversified) batch acquisition styles, considering both BO inspired and Information Retrieval (IR)-inspired variants. 
%anton{considering a BO and am IR style variant for diversification}

\paragraph{Top-B}
This no-diversification strategy selects the top-$B$ documents according to the single-point acquisition function, and is the least computationally expensive since it uses only one computation of $\alpha(z^d|\mathcal{D}^t)$ per batch.

\paragraph{Kriging Believer (KB)}
Starting from a single-point selection, KB \cite{ginsbourger2010kriging} sequentially adds points to $\mathcal{B}^t$ by temporarily assuming the posterior mean \eqref{eqn-gp-pred-pos-mean} as the observation for each selected document, which for uncertainty-aware $\alpha$ such as UCB reduces nearby uncertainty and encourages subsequent selections in other regions. KB requires roughly $B$ times the computation of Top-B since we recompute the posterior $B$ times per batch. 
%After selecting $d^{t+i}$ we add a temporary observation $y^{t+i}=\mu[f(z^{d^{t+i}})]$ to update the posterior before selecting the next document, 

\begin{table}[t]
\footnotesize
\centering
\setlength{\tabcolsep}{4pt}
\caption{BEIR \cite{thakur2021beir} dataset statistics for corpus size, \# of queries, and mean \# of relevant docs.}
\label{tab:datasets}
\begin{tabular}{lrrr}
\toprule
Dataset & \#docs & \#qs & \#rel d/q \\
\midrule
TREC-NEWS & 595K & 57 & 19.6 \\
Robust04 & 528K & 249 & 69.9 \\
TREC-COVID & 171K & 50 & 493.5 \\
SciFact & 5K & 300 & 1.1 \\
NFCorpus & 3.6K & 323 & 38.2 \\

\bottomrule
\end{tabular}
\end{table}

\paragraph{Maximal Marginal Relevance (MMR)}
Inspired by MMR, a classic information retrieval diversity mechanism \cite{carbonell1998use}, we introduce a novel MMR-style document batch acquisition function that sequentially builds $\mathcal{B}^t$ by balancing the value of $\alpha(z^d|\mathcal{D}^t)$ for each $z^d$ with its similarity to already selected documents.
%Starting from $d^{t}=\arg\max_d \alpha(z^d|\mathcal{D}^t)$, we let
% \begin{equation}
% d^{t+i} = \arg\max_{d \in \mathcal{X}\setminus \mathcal{B}^t}
% \left[
% \lambda\,\alpha(z^d|\mathcal{D}^t)
% -
% (1-\lambda)\max_{d' \in \mathcal{B}^t} S(z^d,z^{d'})
% \right],
% \end{equation}
% w
Starting from $d^{t}=\arg\max_d \alpha(z^d|\mathcal{D}^t)$, we let
\begin{equation*}
\begin{aligned}
d^{t+i} = \arg\max_{d \in \mathcal{X}\setminus \mathcal{B}^t}
&\big[\, \lambda\,\alpha(z^d|\mathcal{D}^t) \\
       &-(1-\lambda)\max_{d' \in \mathcal{B}^t} S(z^d,z^{d'}) \big],
\end{aligned}
\end{equation*}
where $S(\cdot,\cdot)$ is a similarity function (e.g., cosine) and $\lambda$ controls the diversity level. Unlike KB, the acquisition function is only computed once, followed by $(B-1)|\mathcal{Z}^{\mathcal{X}}|$ similarity computations which are much cheaper than posterior updates.

\subsection{\method{} Relevance Prediction and Summary}\label{meth:rel-pred}

At any time step $t$, the predicted relevance of a document $d$ is given by the GP posterior mean evaluated at the document embedding,
$\mu[f(z^d)]$ (Eq.~\eqref{eqn-gp-pred-pos-mean}).
This allows \method{} to return ranked documents (with relevance scores) at any stage of the search, making it an anytime retrieval algorithm.
%REPLACE with comment on how \method address top-k retrieval limitations?  
%In summary, \method{} aims to overcome the top-$k$ retrieval stage limitations of unimodal vector similarity retrieval by interleaving high-fidelity LLM query and document relevance observations 

In summary, \method{} initializes a multimodal relevance posterior using the query and its LLM reformulations, and then iteratively acquires (diverse) document batches for LLM query–document scoring and posterior updates, using a BO framework that allows LLM relevance signals to guide the search for relevant documents. We next discuss our experiments to investigate whether \method{} improves on the failures of the top-$k$ retrieval stage of reformulate-retrieve-rerank pipelines.  

\begin{table*}[t]
\footnotesize
\centering
\setlength{\tabcolsep}{3pt}
\caption{Recall(R)@100 and NDCG(N)@10 using Gemini-2.5-Flash-Lite for \method{} with Top-$B$ acquisition ($B=10$) versus LLM rerankers, including IQ (initial query) and QR (query reformulations added) variants. The best overall method is \textbf{bold} while the best absolute LLM scoring method (PW or \method{}) is \underline{underlined}. QR is generally helpful, and \method{} with greedy or UCB acquisition consistently shows large improvements in recall and competitive rankings. }
\label{tab:main}
\begin{tabular}{lcc|cc|cc|cc|cc}
\toprule
\multirow{2}{*}{Method} & \multicolumn{2}{c}{nfcorpus} & \multicolumn{2}{c}{scifact} & \multicolumn{2}{c}{robust} & \multicolumn{2}{c}{covid} & \multicolumn{2}{c}{news} \\
\cmidrule(lr){2-3} \cmidrule(lr){4-5} \cmidrule(lr){6-7} \cmidrule(lr){8-9} \cmidrule(lr){10-11}
 & R@100 & N@10 & R@100 & N@10 & R@100 & N@10 & R@100 & N@10 & R@100 & N@10 \\
\midrule
BM25 Flat & 24.6 & 32.2 & 92.5 & 67.9 & 37.5 & 40.7 & 10.9 & 59.5 & 44.7 & 39.5 \\
BM25 MF & 25.0 & 32.5 & 90.8 & 66.5 & 37.5 & 40.7 & 11.4 & 65.6 & 42.2 & 39.8 \\
SPLADE & 28.4 & 34.7 & 93.5 & 70.4 & 38.5 & 46.8 & 12.8 & 72.7 & 44.1 & 41.5 \\
\midrule
Dense (IQ) & 30.8 & 31.6 & 92.8 & 64.9 & 32.7 & 40.5 & 9.5 & 51.8 & 42.3 & 38.2 \\
PW{} (IQ) & 30.8 & 37.7 & 92.8 & 70.4 & 32.7 & 59.8 & 9.5 & 71.3 & 42.3 & 45.8 \\
\lw{} (IQ) & 30.8 & 36.4 & 92.8 & 73.0 & 32.7 & 51.3 & 9.5 & 74.5 & 42.3 & 44.1 \\
\rankgpt{} (IQ) & 30.8 & 37.6 & 92.8 & 74.4 & 32.7 & 56.7 & 9.5 & \textbf{78.1} & 42.3 & 48.2 \\
\midrule
Dense (QR) & 33.7 & 33.7 & 96.6 & 70.3 & 35.0 & 44.2 & 8.0 & 45.4 & 45.7 & 43.3 \\
PW{} (QR) & 33.7 & 39.3 & 96.6 & 71.0 & 35.0 & 61.2 & 8.0 & 66.5 & 45.7 & 44.0 \\
\lw{} (QR) & 33.7 & 37.3 & 96.6 & 76.1 & 35.0 & 52.8 & 8.0 & 70.6 & 45.7 & 44.0 \\
\rankgpt{} (QR) & 33.7 & 39.9 & 96.6 & \textbf{76.7} & 35.0 & 58.6 & 8.0 & 72.6 & 45.7 & \textbf{49.6} \\
\midrule
\method -IQ-Rand. & 29.9 & 25.9 & 91.2 & 54.5 & 31.9 & 38.2 & 8.0 & 46.7 & 40.3 & 32.8 \\
\method -IQ-Top-$k$ & 34.3 & 38.3 & 92.8 & 66.7 & 38.1 & 59.5 & 13.7 & 75.9 & 45.6 & \underline{45.9} \\
\method -IQ-Greedy & 36.3 & 38.8 & 96.0 & 69.4 & 45.1 & 61.3 & 15.3 & 74.6 & 49.8 & 45.4 \\
\method -IQ-UCB & 35.6 & 38.4 & 93.7 & 69.3 & 44.9 & 60.7 & 15.4 & 70.8 & 52.2 & 42.8 \\
\midrule
\method -QR-Rand. & 32.8 & 33.8 & 96.2 & 70.0 & 36.1 & 47.4 & 12.7 & 64.0 & 47.6 & 44.6 \\
\method -QR-Top-$k$ & 34.3 & 39.6 & 93.5 & \underline{73.3} & 37.6 & 59.6 & 12.9 & 73.0 & 42.4 & 45.8 \\
\method -QR-Greedy & 36.1 & 40.1 & \underline{\textbf{96.7}} & 73.5 & 45.6 & \underline{\textbf{63.6}} & \underline{\textbf{15.6}} & \underline{76.8} & 49.9 & 45.1 \\
\method -QR-UCB & \underline{\textbf{37.7}} & \underline{\textbf{40.2}} & \underline{\textbf{96.7}} & 73.5 & \underline{\textbf{46.5}} & 63.3 & \underline{\textbf{15.6}} & 75.4 & \underline{\textbf{52.7}} & 45.7 \\
\bottomrule
\end{tabular}
\end{table*}

\section{Experimental Setup}
We design a set of experiments to compare \method{} against sparse, dense, and LLM reranker baselines, investigating the effects of LLM query reformulation as well as diversified batched document acquisition and scoring.\footnote{Anonymized code is available at: \url{https://anonymous.4open.science/r/ReBOL-5D56/README.md}} Specifically, we report recall, NDCG, and latency on the five BEIR \cite{thakur2021beir} datasets in Table \ref{tab:datasets} --- these are the five BEIR benchmarks with the fewest queries per dataset (with the exception of Webis-Touche2020 which was found to be highly biased towards token-matching relevance and BM25 in a reproducibility study by the BEIR authors \cite{thakur2024systematic}).
%Our code is available at \anton{Add code link} 

\subsection{\method{} Implementation Details}\label{sec:rebol-impl}
\paragraph{Query Reformulation} Given initial query $q_0$, we prompt Gemini-2.5-Flash-Lite to generate four query reformulations to give a total of five queries with $q_0$, with our exact prompt shown in Fig. \ref{fig:qr-prompt}.

\paragraph{Dense Text Embeddings} As our dense encoder $g:\mathcal{X}\rightarrow \mathcal{Z}^\mathcal{X} \subset \mathbb{R}^m$ we use \texttt{all-MiniLM-L6-v2}\footnote{https://huggingface.co/sentence-transformers/all-MiniLM-L6-v2} with $m = 384$ and normalized embeddings.

\paragraph{Relevance Function and LLM Scoring} We use the 0-3 UMBRELA \cite{upadhyay2024umbrela} relevance scale defined in Section \ref{sec:back-pw} for LLM relevance scoring and to place bounds [0,3] on $f(z)$. 

\paragraph{GP Prior} For our GP prior, we use a 0 mean and RBF kernel \eqref{eqn-rbf} with $\sigma_s$, $\sigma_n$, and $\ell$ set to 1. 

\paragraph{Acquisition Functions} As per Section \ref{sec:meth-af}, and using a batch size $B=10$ unless otherwise stated, we test three batch acquisition mechanisms: Top-$B$, KB, and MMR with $\lambda \in \{0.5,0.7,0.9\}$. These batch functions rely on some single-point acquisition function, for which we test greedy, random, and UCB with $\beta$ set to 1. 

\paragraph{Number of Observations} In our main experiments, \method{}  observes exactly 100 documents iteratively selected by the acquisition function, while the appendix includes ablation studies with 50 and 200 observations.   

%and LLM baselines observe exactly 100 distinct documents, which for \method{} are selected by the acquisition function, and for LLM rerankers are the top-100 documents from dense retrieval. The appendix includes ablation studies with 50 and 200 observations.   

\subsection{Baseline Implementations}
\subsubsection{Vector Similarity Retrieval}

\begin{table*}[t]
\footnotesize
\centering
\setlength{\tabcolsep}{6pt}
\caption{Batch size $B=1$ vs. 10 ablation for \method-QR (Top-$B$) versus PW Reranking (QR) showing Recall(R)@100 and NDCG(N)@10, as well as mean time per query spent in LLM calls versus in total --- both PW and \method{} see large latency reductions due to batching. 
%PW performance is comparable for both $B=1$ and 10, while 
\method{} performs noticeably better with $B=1$, showing the benefits of more interleaved LLM scoring and BO if time is available.}
\label{tab:batch-runtime}

\begin{tabular}{l c cc|cc || cc|cc}
\toprule
\multirow{3}{*}{Method} & \multirow{3}{*}{Batch}
& \multicolumn{4}{c}{covid}
& \multicolumn{4}{c}{news} \\
\cmidrule(lr){3-6}
\cmidrule(lr){7-10}

&
& \multicolumn{2}{c}{Metric}
& \multicolumn{2}{c}{Time (s)}
& \multicolumn{2}{c}{Metric}
& \multicolumn{2}{c}{Time (s)} \\
\cmidrule(lr){3-4}
\cmidrule(lr){5-6}
\cmidrule(lr){7-8}
\cmidrule(lr){9-10}

&
& R@100 & N@10 & LLM & Total
& R@100 & N@10 & LLM & Total \\
\midrule

Dense (QR) & n/a
& 8.0 & 45.4 & 0.8 & 1.0
& 45.7 & 43.3 & 0.6 & 1.1 \\

\midrule

PW (QR) & \multirow{3}{*}{1}
& 8.0 & 65.9 & 38.2 & 38.4
& 45.9 & 44.7 & 37.3 & 38.0 \\

\method{}-QR-Greedy &
& 15.6 & 79.6 & 39.4 & 42.9
& 51.3 & \textbf{\underline{48.4}} & 37.6 & 41.9 \\

\method{}-QR-UCB &
& \textbf{\underline{16.4}} & \textbf{\underline{82.4}} & 39.4 & 46.9
& \textbf{\underline{52.8}} & 47.9 & 39.3 & 63.7 \\

\midrule

PW (QR) & \multirow{3}{*}{10}
& 8.0 & 66.5 & \textbf{\underline{4.9}} & \textbf{\underline{5.0}}
& 45.9 & 44.0 & \textbf{\underline{5.5}} & \textbf{\underline{5.9}} \\

\method{}-QR-Greedy &
& 15.6 & 76.8 & 5.4 & 5.9
& 49.9 & 45.1 & 5.7 & 6.3 \\

\method{}-QR-UCB &
& 15.6 & 75.4 & 5.4 & 6.4
& 52.7 & 45.7 & 5.8 & 8.4 \\

\bottomrule
\end{tabular}
\end{table*}

\paragraph{Dense Retrieval} Our dense retrieval baseline uses the same text encoder as in Section \ref{sec:rebol-impl} and dot product similarity. We also test dense retrieval with MMR \cite{carbonell1998use} at the same $\lambda$ values as for \method-MMR in Sec. \ref{sec:rebol-impl}.  

\paragraph{Dense Retrieval (QR)} Our LLM query reformulation dense retrieval baseline uses the same query generation process as for \method{} in Sec. \ref{sec:rebol-impl}. We then follow the retrieval process shown in Fig. \ref{fig:qr}a, in which $Q+1$ similarity scores (one for each reformulation plus $q_0$) are computed using dense retrieval then averaged for the final ranking. 

\paragraph{Sparse Retrieval} We report several \texttt{pyserini}\footnote{https://github.com/castorini/pyserini} sparse retrieval baselines including BM25-Flat, BM25 Multi-Field (MF), and SPLADE. 

\subsubsection{LLM Reranking} All LLM rerankers judge the top-$k$ documents from dense retrieval, with $k$ fixed to 100 in our main experiments, matching the number of observations made by \method{}. The appendix includes ablations with $k = 50$ and $k = 200$. 

\paragraph{LW LLM Reranking} As per Section \ref{sec:back-lw}, we test two LW variants. The first, \textit{\lw} reranks all $k$ documents in a single LLM call, using the prompt in Fig \ref{fig:lw-prompt}. The second, \textit{\rankgpt}, uses the same prompt but with an upwards sliding window of $W$ documents --- for which we use a step size $\frac{W}{2}$ and $W=20$ to maintain a comparable number of LLM calls (i.e., 9) for $k=100$ to \method{} and PW reranking (10 LLM calls with batch size 10). 
  
\paragraph{(Batched) PW LLM Reranking} PW LLM reranking uses the same rubric (Sec. \ref{sec:back-pw}) as \method{}, and unless otherwise stated, also uses a default batch size of $B=10$. 

\subsubsection{\method{}-Top-$k$ Ablation}\label{sec:exp:top-k} We include a version of \method{} which, after initializing the posterior with query observations, simply makes observations at the dense retriever top-$k$ instead of iteratively using an acquisition function, allowing us to test whether active acquisition is beneficial.

\section{Experimental Results}
We first report recall and ranking performance of \method{} vs. baselines, testing both IQ and QR variants with the simplest top-$B$ batch acquisition method (Tab. \ref{tab:main}). We then look at performance and latency results for different batch sizes (1 vs. 10) and various batch acquisition strategies (Table \ref{tab:batch-runtime}-\ref{tab:main_mean}).

%All results in the main paper use Gemini-2.5-Flash-Lite and $k=100$, with ablations on the LLM scoring model (GPT-5.2, cross-encoder) and $k \in \{50,200\}$ shown in the appendix.  

\begin{table*}[t]
\footnotesize
\centering
\setlength{\tabcolsep}{4pt}
\caption{Effect of diversification methods, including MMR ($x=\lambda$), Kriging Believer, and Top-$B$ (no diversity).}
\label{tab:batch-and-div}

\begin{tabular}{l l cc|cc|cc|cc|cc}
\toprule
\multirow{2}{*}{Method} & \multirow{2}{*}{Diversity}
& \multicolumn{2}{c}{nfcorpus}
& \multicolumn{2}{c}{scifact}
& \multicolumn{2}{c}{robust}
& \multicolumn{2}{c}{covid}
& \multicolumn{2}{c}{news} \\

\cmidrule(lr){3-4}
\cmidrule(lr){5-6}
\cmidrule(lr){7-8}
\cmidrule(lr){9-10}
\cmidrule(lr){11-12}

&
& \multicolumn{1}{c}{R@100} & \multicolumn{1}{c}{N@10}
& \multicolumn{1}{c}{R@100} & \multicolumn{1}{c}{N@10}
& \multicolumn{1}{c}{R@100} & \multicolumn{1}{c}{N@10}
& \multicolumn{1}{c}{R@100} & \multicolumn{1}{c}{N@10}
& \multicolumn{1}{c}{R@100} & \multicolumn{1}{c}{N@10} \\

\midrule

\multirow{4}{*}{Dense (QR)}
& n/a
& 30.8 & 31.6
& 92.8 & 64.9
& 32.7 & 40.5
& 9.5 & 51.8
& 42.3 & 38.2 \\

& MMR 0.9
& 33.0 & 33.6
& 96.6 & 70.9
& 34.4 & 43.9
& 8.1 & 46.2
& 44.2 & 41.7 \\

& MMR 0.7
& 29.2 & 30.8
& 94.9 & 71.1
& 30.0 & 40.7
& 7.8 & 41.6
& 36.0 & 34.8 \\

& MMR 0.5
& 15.6 & 21.9
& 74.5 & 63.7
& 9.7 & 24.7
& 1.9 & 26.7
& 8.2 & 21.9 \\

\midrule

\multirow{5}{*}{\method{}-QR-Greedy}
& Top-B
& 36.1 & 40.1
& \textbf{96.7} & 73.5
& 45.6 & 63.6
& 15.6 & 76.8
& 49.9 & 45.1 \\

% & KB
% & 36.2 & 40.3
% & 96.0 & 74.4
% & 43.9 & 63.2
% & 15.1 & 77.5
% & 50.6 & 44.9 \\

& MMR 0.9
& \textbf{38.6} & 41.3
& 96.3 & 73.5
& 46.1 & 63.9
& 15.9 & 80.5
& 50.5 & 47.4 \\

& MMR 0.7
& 38.5 & \textbf{41.6}
& 96.0 & 74.2
& 46.8 & 63.8
& 15.9 & 78.7
& 51.4 & 45.8 \\

& MMR 0.5
& 37.4 & 40.8
& 96.0 & 75.0
& 45.8 & 64.2
& 16.0 & 81.7
& 52.2 & 47.2 \\

\midrule

\multirow{5}{*}{\method{}-QR-UCB}
& Top-B
& 37.7 & 40.2
& \textbf{96.7} & 73.5
& 46.5 & 63.3
& 15.6 & 75.4
& 52.7 & 45.7 \\

& KB
& 37.6 & 40.7
& 96.5 & 72.2
& 45.6 & 63.7
& 16.0 & \textbf{83.2}
& 51.7 & 45.5 \\

& MMR 0.9
& 37.9 & 41.1
& 95.7 & 73.0
& 46.7 & 64.6
& 15.9 & 77.3
& 52.4 & \textbf{49.2} \\

& MMR 0.7
& 37.9 & 41.0
& 96.2 & 72.3
& \textbf{47.1} & \textbf{65.2}
& \textbf{16.1} & 78.2
& \textbf{54.2} & 47.5 \\

& MMR 0.5
& 38.3 & 41.5
& 95.7 & \textbf{75.2}
& 44.7 & 63.8
& \textbf{16.1} & \textbf{83.2}
& 52.2 & 48.2 \\

\bottomrule
\end{tabular}
\end{table*}

\begin{table*}[t]
\footnotesize
\centering
\setlength{\tabcolsep}{6pt}

\caption{Mean time (s) per query spent in LLM calls versus in total, with $B=10$ for \method{} and PW.}
\label{tab:main_mean}

\begin{tabular}{l l cc|cc|cc|cc|cc}
\toprule
Method & Diversity 
& \multicolumn{2}{c}{nfcorpus}
& \multicolumn{2}{c}{scifact}
& \multicolumn{2}{c}{robust}
& \multicolumn{2}{c}{covid}
& \multicolumn{2}{c}{news} \\

\cmidrule(lr){3-4}
\cmidrule(lr){5-6}
\cmidrule(lr){7-8}
\cmidrule(lr){9-10}
\cmidrule(lr){11-12}

& 
& \multicolumn{1}{c}{LLM} & \multicolumn{1}{c}{Total}
& \multicolumn{1}{c}{LLM} & \multicolumn{1}{c}{Total}
& \multicolumn{1}{c}{LLM} & \multicolumn{1}{c}{Total}
& \multicolumn{1}{c}{LLM} & \multicolumn{1}{c}{Total}
& \multicolumn{1}{c}{LLM} & \multicolumn{1}{c}{Total} \\

\midrule

Dense & n/a
& 0.0 & 0.1
& 0.0 & 0.1
& 0.0 & 0.4
& 0.0 & 0.2
& 0.0 & 0.4 \\

\midrule

Dense (QR) & n/a
& 0.9 & 0.9
& 0.8 & 0.9
& 1.1 & 1.4
& 0.8 & 1.0
& 0.6 & 1.1 \\

PW (QR) & n/a
& 5.0 & 5.0
& 5.0 & 5.0
& 5.6 & 6.0
& 4.9 & 5.0
& 5.5 & 5.9 \\

\lw{} (QR) & n/a
& 4.1 & 4.1
& 4.3 & 4.3
& 4.4 & 4.8
& 2.7 & 2.8
& 4.5 & 5.0 \\

\rankgpt{} (QR) & n/a
& 8.3 & 8.3
& 9.7 & 9.8
& 11.9 & 12.3
& 7.3 & 7.4
& 15.7 & 16.7 \\

\midrule

\method{}-QR-Rand. & Top-B
& 6.0 & 6.1
& 7.1 & 7.3
& 5.7 & 5.9
& 5.2 & 5.6
& 5.7 & 6.0 \\

\method{}-QR-Top-$k$ & n/a
& 5.7 & 5.8
& 6.3 & 6.4
& 6.0 & 6.2
& 5.9 & 6.0
& 5.7 & 5.9 \\

\midrule

\method{}-QR-Greedy & Top-B
& 5.9 & 6.2
& 5.8 & 6.1
& 6.2 & 6.7
& 5.4 & 5.9
& 5.7 & 6.3 \\

\method{}-QR-UCB & Top-B
& 6.0 & 6.3
& 7.2 & 7.6
& 5.9 & 8.2
& 5.4 & 6.4
& 5.8 & 8.4 \\

\midrule

% \method{}-QR-Greedy & KB
% & 5.1 & 6.1
% & 4.9 & 5.9
% & 8.2 & 12.0
% & 6.3 & 8.9
% & 6.4 & 10.3 \\

\method{}-QR-UCB & KB
& 5.1 & 6.1
& 4.8 & 5.9
& 8.1 & 29.2
& 7.5 & 14.7
& 7.5 & 31.1 \\

\midrule

\method{}-QR-Greedy & MMR
& 5.4 & 5.9
& 5.9 & 6.6
& 6.4 & 13.3
& 6.4 & 8.8
& 6.7 & 14.3 \\

\method{}-QR-UCB & MMR
& 5.4 & 5.9
& 5.1 & 5.7
& 5.9 & 14.4
& 7.1 & 10.0
& 8.6 & 18.4 \\

\bottomrule
\end{tabular}
\end{table*}

%GPT: tab:gpt52_ce
%50 obs: tab:50-obs-ablation
%200 obs: tab:200-obs-ablation

\paragraph{RQ1: How does \method{} performance compare against LLM reranker baselines?}
\method{} with greedy and UCB Top-$B$ acquisition achieves large improvements in recall and competitive NDCG compared to LLM rerankers in Table \ref{tab:main}, as well as in ablation studies using different LLM scoring models (GPT-5.2, cross-encoder) and numbers of LLM observations (App. Tables \ref{tab:gpt52_ce}-\ref{tab:200-obs-ablation}). Further, comparison to the \method{}-Top-$k$ ablation (Sec. \ref{sec:exp:top-k}) shows that active learning improves recall considerably. 
\paragraph{RQ2: How helpful is LLM query reformulation for retrieval and \method?}
As shown in Table \ref{tab:main}, QR is helpful for both LLM rerankers and \method{}, improving recall and NDCG for all datasets except TREC-COVID --- where QR is detrimental for LLM rerankers but helpful for \method{}, suggesting that \method{} is more robust in adapting to different relevance distributions due to BO. Further, QR shows warm start benefits, as indicated by the strong lift of QR on \method{}-Random, and the ablation with 50 observations (Appendix Table \ref{tab:50-obs-ablation}) showing greater QR improvements for \method{} than for the default 100 observations.   
%for \method{} --- it is particularly helpful for NDCG@10 indicating that the query reformulations may especially help to strengthen high relevance peaks. 
\paragraph{RQ3: What are the effects of various batching strategies on performance?}
The ablation in Table \ref{tab:batch-runtime} shows that $B=1$ can do well for \method{}-QR since it uses the most recent information for each acquisition, but requires a long time per query in LLM calls (e.g., 36s$+$), while a batch size of $B=10$ reduces latency significantly (e.g., 5s). Table \ref{tab:batch-and-div} thus compares different \method{} batch diversification strategies, showing that MMR typically performs best. MMR also helps dense retrieval, but its benefits are amplified by the BO framework of \method{}. 
\paragraph{RQ4: How does \method{} compare against LLM rerankers in terms of latency?}
Table \ref{tab:main_mean} shows that \method{} with Top-$B$ acquisition does not add much latency compared to LLM rerankers since LLM time dominates. MMR increases latency proportionally to corpus size as it requires roughly ($k-B$)$|\mathcal{X}|$ similarity evaluations, but is the best performing batching method in Table \ref{tab:batch-and-div}. KB takes the longest but doesn't generally improve performance.  

\section{Conclusion} This work addresses the limitations of the top-$k$ retrieval stage of LLM reformulate-retrieve-rerank pipelines by proposing \method{}. \method{} reframes retrieval as Bayesian Optimization (BO) and first initializes a multimodal posterior with several LLM query reformulations (QR). It then iteratively uses a (diversified) batch acquisition function to select documents, generate LLM query-document relevance scores, and update the BO posterior to optimize relevance. After introducing new techniques for BO retrieval via LLM-QR, batched LLM relevance judgments, and MMR-style batch diversification, we conduct a range of experiments comparing \method{} to LLM rerankers and vector retrieval baselines on five datasets. Our experiments indicate that \method{} is a promising new paradigm for improving retrieval and can effectively leverage query reformulations and diversified batch acquisition.    

\section{Limitations}
Our work includes the following limitations. Firstly, we only perform experiments on five datasets, since LLM experiments are very computationally expensive and our tests require testing a number of configurations, including reranking variants and different acquisition functions for our method. In this regard, another limitation is that we only evaluate two LLMs which are Gemini-2.5-Flash-Lite and GPT-5.2 as well as a cross-encoder (ms-marco-MiniLM-L6-v2) as scoring functions for \method{} and pointwise reranking. We also only report results for three levels of $k$ (i.e. LLM observation number), specifically 50,100, and 200. 

Several broader risks also arise when using LLMs for large-scale ranking and relevance assessment. First, LLMs may reproduce or amplify societal biases learned during pretraining, which can lead to harmful or unfair retrieval outcomes. Second, LLM-based systems remain vulnerable to adversarial prompt manipulation (e.g., prompt injection or “jailbreaking”), which can compromise system safety or reliability. Third, LLM relevance judgments themselves may be incorrect or inconsistent, particularly in ambiguous or domain-specific cases. Such errors may propagate through retrieval pipelines and could be problematic in high-stakes applications where ranking quality is critical. 

As per ARR guidelines, we disclose the use of AI assistants for assistance with coding and grammar.

\bibliography{custom}

\FloatBarrier
\appendix
\section{A. Prompts}
\begin{figure}
    \centering
    \includegraphics[width=0.75\linewidth]{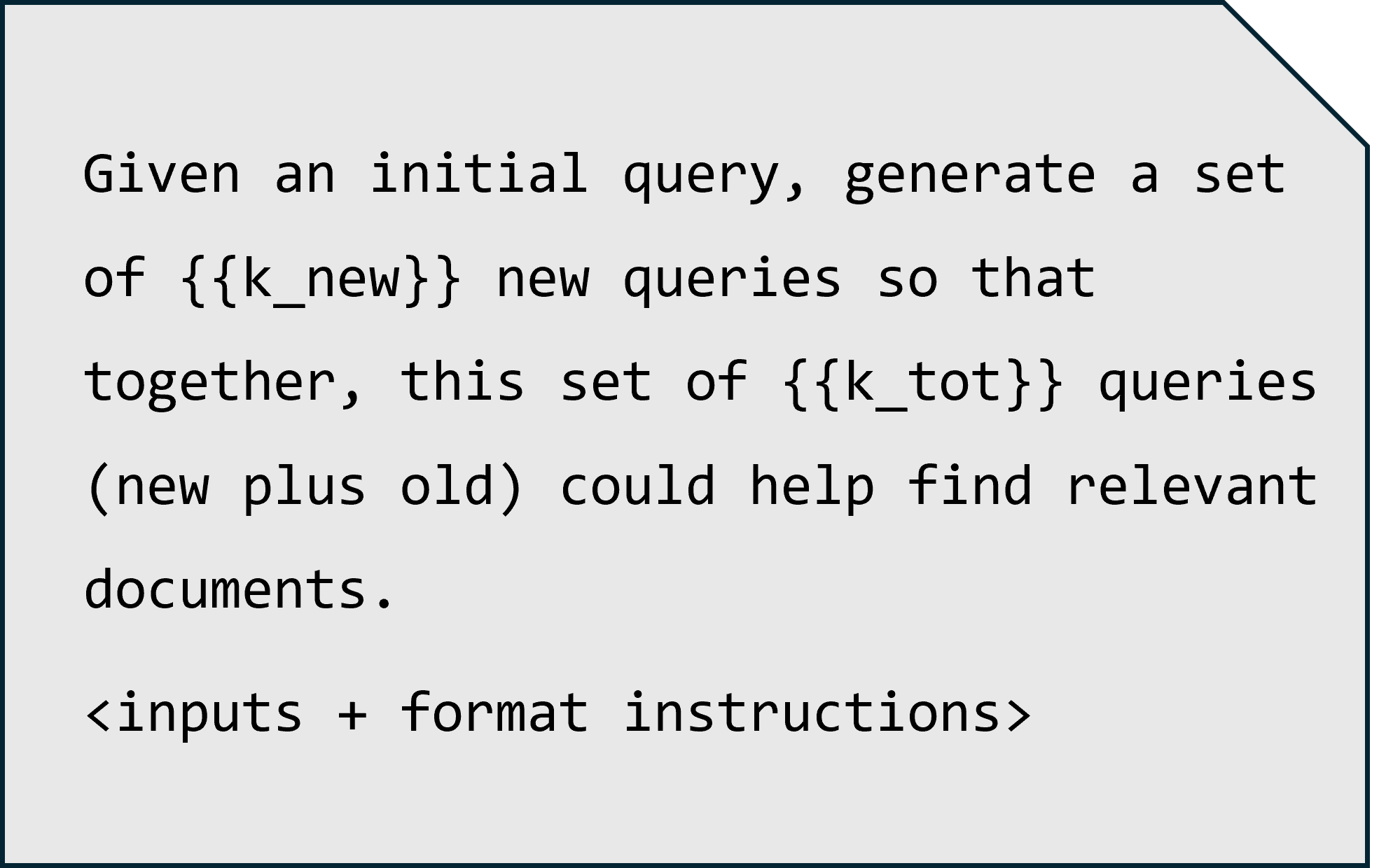}
    \caption{Query Reformulation Prompt}
    \label{fig:qr-prompt}
\end{figure}

\begin{figure}
    \centering
    \includegraphics[width=0.75\linewidth]{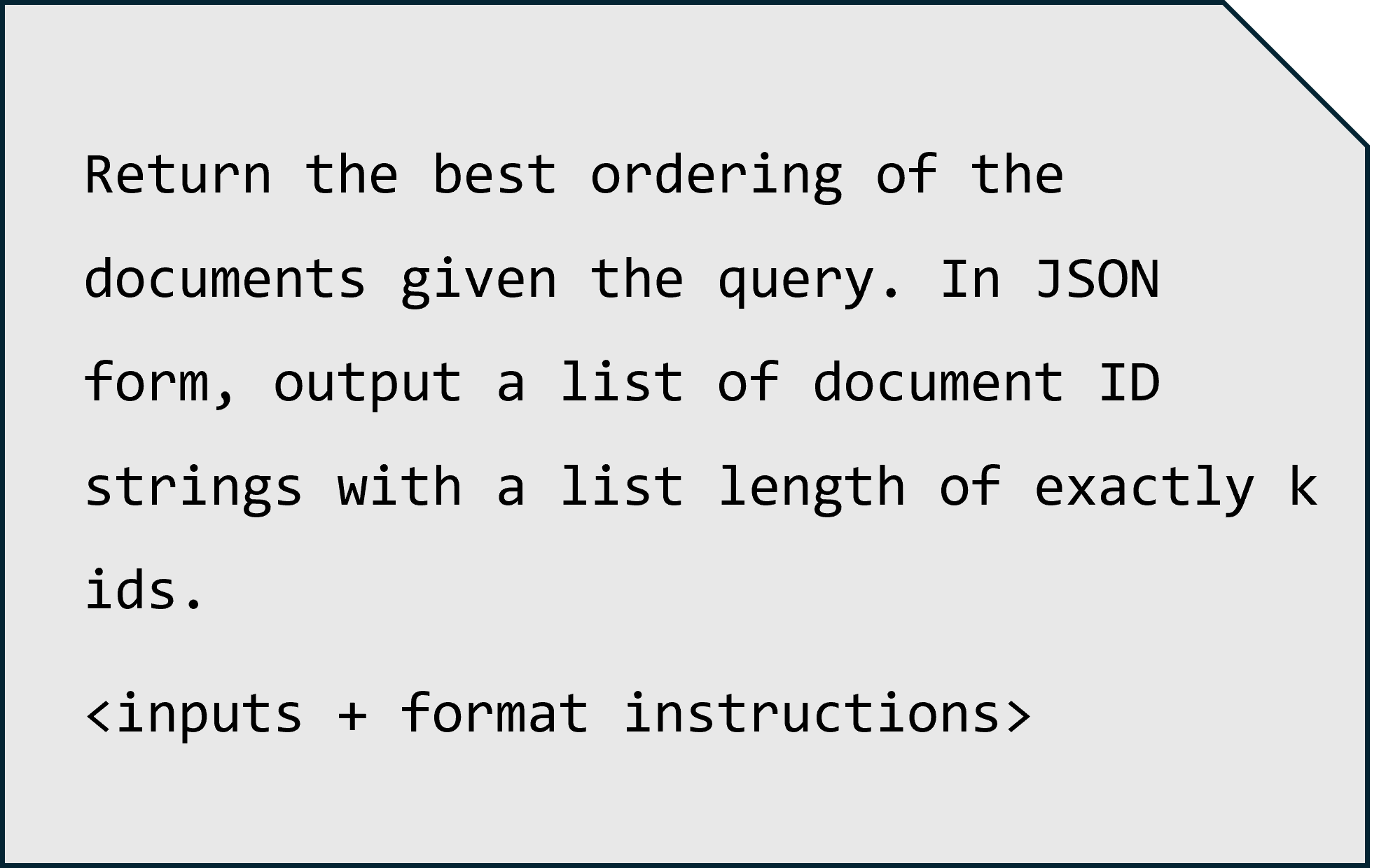}
    \caption{Listwise reranking prompt --- real document IDs are simplified before prompting to ["d1" ,$...$, "dk"] for $k$ documents.}
    \label{fig:lw-prompt}
\end{figure}

Figure \ref{fig:qr-prompt}-\ref{fig:lw-prompt} show our prompts for query reformulation and listwise reranking. 

\section{B. Ablations}
%appendix plan
% \begin{itemize}
%     \item Prompts
%     \item @X NDCG and Recall charts
%     \item 50 and 100 observation ablations
%     \item CE full results
%     \item Time breakdowns
% \end{itemize}

\begin{table*}
\small
\centering
\caption{LLM relevance observation model ablation with GPT-5.2 and a cross-encoder (ms-marco-MiniLM-L6-v2) showing Recall(R)@100 and NDCG(N)@10 for \method{} (Top-$B$), with \textbf{bold} indicating the best overall method and \underline{underline} the best absolute relevance scorer. In general, \method{} continues to display strong performance though it is relatively weaker with the cross-encoder.}

\label{tab:gpt52_ce}
\begin{tabular}{l|cc|cc||cc|cc}
\toprule
\multirow{3}{*}{Method} 
& \multicolumn{4}{c}{GPT-5.2} 
& \multicolumn{4}{c}{Cross-Encoder} \\
\cmidrule(lr){2-5} \cmidrule(lr){6-9}
& \multicolumn{2}{c}{covid} & \multicolumn{2}{c}{news}
& \multicolumn{2}{c}{covid} & \multicolumn{2}{c}{news} \\
\cmidrule(lr){2-3} \cmidrule(lr){4-5}
\cmidrule(lr){6-7} \cmidrule(lr){8-9}
& R@100 & N@10 & R@100 & N@10 & R@100 & N@10 & R@100 & N@10 \\
\midrule
Dense (IQ) & 9.5 & 51.8 & 42.3 & 38.2 & 9.5 & 51.8 & 42.3 & 38.2 \\
PW (IQ) & 9.5 & 73.8 & 42.3 & 45.9 & 9.5 & \textbf{\underline{74.0}} & 42.3 & 45.6 \\
\rankgpt{} (IQ) & 9.5 & 78.0 & 42.3 & 52.4 & n/a & n/a & n/a & n/a \\
\midrule
Dense (QR) & 8.0 & 45.4 & 45.7 & 43.3 & 8.0 & 45.4 & 45.7 & 43.3 \\
PW (QR) & 8.0 & 63.5 & 45.7 & 47.5 & 8.0 & 67.4 & 45.7 & 46.5 \\
\rankgpt{} (QR) & 8.0 & 66.7 & 45.7 & \textbf{53.3} & n/a & n/a & n/a & n/a \\
\midrule
%\method -Top-$k$ & 12.9 & 76.8 & 42.8 & 42.0 & 12.3 & 68.0 & 35.1 & 45.2 \\
\method -IQ-Greedy & 15.0 & 79.7 & \textbf{\underline{55.0}} & 45.9 & 12.7 & 70.4 & 47.0 & 47.4 \\
\method -IQ-UCB & 14.9 & 76.8 & 54.2 & 47.5 & 13.0 & 70.2 & 47.8 & 47.1 \\
\midrule
\method -QR-Greedy & 15.2 & \textbf{\underline{80.8}} & 54.0 & \underline{50.8} & \textbf{\underline{13.5}} & 73.8 & 47.2 & \textbf{\underline{49.3}} \\
\method -QR-UCB & \textbf{\underline{15.4}} & 80.2 & 53.1 & 47.9 & 12.4 & 65.8 & \textbf{\underline{48.6}} & 49.2 \\
\bottomrule
\end{tabular}
\end{table*}

\begin{table*}[t]
\footnotesize
\centering
\caption{Ablation study with $k = 50$ and 50 observations for \method{} (Top $B$)}
\label{tab:50-obs-ablation}

\begin{tabular}{l cc|cc|cc|cc|cc}
\toprule
Method
& \multicolumn{2}{c}{nfcorpus}
& \multicolumn{2}{c}{scifact}
& \multicolumn{2}{c}{robust}
& \multicolumn{2}{c}{covid}
& \multicolumn{2}{c}{news} \\
\cmidrule(lr){2-3}\cmidrule(lr){4-5}\cmidrule(lr){6-7}\cmidrule(lr){8-9}\cmidrule(lr){10-11}
& R@50 & N@10
& R@50 & N@10
& R@50 & N@10
& R@50 & N@10
& R@50 & N@10 \\
\midrule

Dense (IQ)
& 25.6 & 31.6
& 87.9 & 64.9
& 25.7 & 40.5
& 5.6 & 51.8
& 33.8 & 38.2 \\

PW (IQ)
& 25.6 & 37.2
& 87.9 & 70.3
& 25.7 & 57.2
& 5.6 & 70.6
& 33.8 & 47.4 \\

\lw{} (IQ)
& 25.6 & 35.7
& 87.9 & 72.1
& 25.7 & 52.2
& 5.6 & 70.3
& 33.8 & 46.7 \\

\rankgpt{} (IQ)
& 25.6 & 37.1
& 87.9 & 72.8
& 25.7 & 55.7
& 5.6 & 75.1
& 33.8 & 49.1 \\

\midrule

Dense (QR)
& 28.0 & 33.7
& 92.9 & 70.3
& 28.0 & 44.2
& 5.0 & 45.4
& 36.8 & 43.3 \\

PW (QR)
& 28.0 & 39.3
& 92.9 & 72.3
& 28.0 & 59.6
& 5.0 & 62.3
& 36.8 & 46.4 \\

\lw{} (QR)
& 28.2 & 39.1
& 92.9 & 75.7
& 28.0 & 54.2
& 5.0 & 62.7
& 36.8 & 47.1 \\

\rankgpt{} (QR)
& 28.2 & 40.2
& 92.9 & \textbf{76.6}
& 28.0 & 57.6
& 5.0 & 64.5
& 36.8 & 48.1 \\

\midrule

\method{}-Rand.(IQ)
& 23.8 & 24.9
& 87.5 & 57.7
& 26.0 & 39.1
& 5.7 & 47.1
& 33.6 & 37.8 \\

\method{}-Top-$k$ (IQ)
& 29.4 & 38.3
& 92.1 & 66.7
& 32.5 & 59.5
& 8.1 & \underline{\textbf{75.9}}
& 38.1 & 45.9 \\

\method{}-Greedy (IQ)
& 28.7 & 38.9
& 91.5 & 70.2
& 34.1 & 59.1
& 8.2 & 73.4
& 37.7 & 47.5 \\

\method{}-UCB (IQ)
& 29.2 & 37.9
& 90.9 & 69.2
& 33.6 & 58.3
& \underline{\textbf{8.6}} & 72.4
& 39.7 & 45.5 \\

\midrule

\method{}-Rand. (QR)
& 26.7 & 32.6
& 92.6 & 70.4
& 29.4 & 47.1
& 6.8 & 56.8
& 38.1 & 45.3 \\

\method{}-Top-$k$ (QR)
& 29.9 & 39.6
& 93.0 & 73.3
& 32.3 & 59.6
& 7.7 & 73.0
& 34.1 & 45.8 \\

\method{}-Greedy (QR)
& 30.3 & 40.0
& \underline{\textbf{94.0}} & 73.1
& 35.1 & \underline{\textbf{62.9}}
& \underline{\textbf{8.6}} & 75.4
& 37.0 & 44.3 \\

\method{}-UCB (QR)
& \underline{\textbf{31.9}} & \underline{\textbf{40.9}}
& 93.7 & \underline{73.9}
& \underline{\textbf{35.4}} & 60.8
& 8.4 & 74.2
& \underline{\textbf{40.6}} & \underline{\textbf{49.2}} \\

\bottomrule
\end{tabular}
\end{table*}

\begin{table*}[t]
\footnotesize
\centering
\caption{Ablation study with $k=200$ and 200 observations for \method{}  (Top $B$)}
\label{tab:200-obs-ablation}

\begin{tabular}{l cc|cc|cc|cc|cc}
\toprule
Method
& \multicolumn{2}{c}{nfcorpus}
& \multicolumn{2}{c}{scifact}
& \multicolumn{2}{c}{robust}
& \multicolumn{2}{c}{covid}
& \multicolumn{2}{c}{news} \\
\cmidrule(lr){2-3}\cmidrule(lr){4-5}\cmidrule(lr){6-7}\cmidrule(lr){8-9}\cmidrule(lr){10-11}
& R@100 & N@10
& R@100 & N@10
& R@100 & N@10
& R@100 & N@10
& R@100 & N@10 \\
\midrule

Dense (IQ)
& 30.8 & 31.6
& 92.8 & 64.9
& 32.7 & 40.5
& 9.5 & 51.8
& 42.3 & 38.2 \\

PW (IQ)
& 32.3 & 38.3
& 95.2 & 69.9
& 38.0 & 61.0
& 11.6 & 72.0
& 46.7 & 44.8 \\

\lw{} (IQ)
& 30.7 & 33.6
& 93.1 & 69.2
& 33.3 & 48.7
& 10.1 & 72.8
& 42.3 & 41.4 \\

\rankgpt{} (IQ)
& 32.3 & 37.3
& 95.2 & 76.0
& 35.8 & 58.5
& 10.4 & \textbf{81.5}
& 44.6 & 48.3 \\

\midrule

Dense (QR)
& 33.3 & 34.7
& 95.8 & 71.0
& 35.0 & 44.4
& 8.3 & 46.3
& 46.1 & 42.2 \\

PW (QR)
& 34.9 & 38.5
& 97.0 & 68.3
& 40.5 & 60.8
& 10.3 & 66.3
& 49.8 & 43.5 \\

\lw{} (QR)
& 33.5 & 36.2
& 95.5 & 72.2
& 35.1 & 49.8
& 9.1 & 70.6
& 45.8 & 43.5 \\

\rankgpt{} (QR)
& 34.8 & 40.5
& 97.0 & \textbf{77.1}
& 38.1 & 59.3
& 9.1 & 77.4
& 48.4 & \textbf{50.0} \\

\midrule

\method{}-Rand (IQ).
& 30.8 & 25.3
& 90.7 & 49.5
& 31.9 & 38.2
& 7.3 & 39.5
& 40.8 & 36.1 \\

\method{}-Top-$k$ (IQ)
& 36.9 & 39.1
& 95.4 & 66.5
& 40.8 & 60.4
& 14.0 & 77.5
& 46.8 & 43.0 \\

\method{}-Greedy (IQ)
& 38.0 & 39.1
& 96.3 & 69.3
& 49.5 & 61.6
& 16.8 & 75.0
& 53.8 & 45.0 \\

\method{}-UCB (IQ)
& 37.8 & 37.9
& 95.0 & 68.5
& 49.1 & 61.5
& 16.3 & 75.9
& \underline{\textbf{55.8}} & 40.4 \\

\midrule

\method{}-Rand. (QR)
& 34.3 & 33.4
& 95.8 & 69.5
& 36.1 & 47.2
& 12.6 & 62.3
& 46.7 & 43.8 \\

\method{}-Top-$k$ (QR)
& 36.3 & 40.2
& 96.5 & 73.9
& 41.1 & 61.0
& 13.4 & 74.6
& 46.6 & 44.8 \\

\method{}-Greedy (QR)
& 38.6 & 40.1
& \underline{\textbf{98.7}} & \underline{74.1}
& \underline{\textbf{50.3}} & 63.5
& 16.8 & 75.8
& 53.3 & 45.7 \\

\method{}-UCB (QR)
& \underline{\textbf{39.2}} & \underline{\textbf{40.7}}
& 96.0 & 73.6
& 50.0 & \underline{\textbf{64.3}}
& \underline{\textbf{17.2}} & 77.0
& 54.8 & \underline{47.1} \\

\bottomrule
\end{tabular}
\end{table*}

%\section{Example Appendix}
%\label{sec:appendix}

%This is an appendix.

\end{document}